# A Cut-Based BAT-MCS Approach for Binary-State Network Reliability Assessment

Wei-Chang Yeh
Department of Industrial Engineering and
Engineering Management
National Tsing Hua University, Hsinchu, Taiwan
yeh@ieee.org

*Abstract* — The BAT-MCS is an integrated Monte Carlo simulation method (MCS) that combines a binary adaptation tree algorithm (BAT) with a self-regulating simulation mechanism. The BAT algorithm operates deterministically, while the Monte Carlo simulation method is stochastic. By hybridizing these two approaches, BAT-MCS successfully reduces variance, increases efficiency, and improves the quality of its binary-state network reliability. However, it has two notable weaknesses. First, the selection of the supervectors, sub-vectors that form the core of BAT-MCS, is overly simplistic, potentially affecting overall performance. Second, the calculation of the approximate reliability is complicated, which limits its strength in reducing variance. In this study, a new BAT-MCS called cBAT-MCS is proposed to enhance the performance of the BAT-MCS. The approach reduces the complexity of MCS. Selecting the super-vector based on a novel layer-cut approach can reduce both runtime and variance. Extensive numerical experiments on large-scale binary-state network demonstrate that the proposed new cBAT-MCS outperforms traditional MCS and original BAT-MCS approaches in terms of computational efficiency and accuracy.

Keywords: Binary-State Network Reliability, Monte Carlo Simulation (MCS); Binary-Addition-Tree algorithm (BAT); BAT-MCS; Supervector

## 1. INTRODUCTION

Binary-state networks serve as fundamental tools in reliability analysis and network management by representing each component in one of two states—operational or failed [1]. This simplified yet powerful framework enables precise reliability calculations and systematic assessments of complex systems [1], making it invaluable for identifying vulnerabilities and optimizing maintenance strategies in critical infrastructure management [2]. The elegance of binary representation lies in its ability to model intricate system behaviors through discrete state transitions, thereby providing a solid mathematical foundation for reliability engineering and risk assessment [1].

Moreover, the applications of binary-state networks span diverse domains [40]—including network resilience [3], wireless sensor networks [4], pipe networks [5], Internet of Things systems [6], data theft attack analysis, social networks [7], transportation networks [36], and Markovian systems [8] — with platforms like social networks now encompassing over 3.6 billion users [7]. Such exponential growth in network size and complexity presents unprecedented challenges in reliability assessment and system optimization.

Network reliability—defined as the probability of successful network functioning—serves as a crucial performance metric, directly impacting system availability [26, 27, 30, 32, 37], maintenance scheduling [31, 33], and resource allocation decisions [9, 25, 35]. However, calculating exact binary-state network reliability is an NP-hard problem [10, 28, 34], necessitating the use of Monte Carlo simulation (MCS) methods for reliable estimation across various network types [11, 29, 38].

While MCS methods have proven invaluable in analyzing complex systems where analytical solutions are impractical, traditional approaches often face challenges with computational cost and variance, particularly in large-scale binary-state networks [12]. These limitations become especially pronounced when analyzing networks with high component counts or complex interdependencies [13]. To address these limitations, researchers have developed hybrid methods that combine deterministic and stochastic strategies, leveraging the strengths of both approaches [14].

To mitigate these challenges, researchers have developed hybrid methods that combine deterministic and stochastic strategies. One notable advancement is the BAT-MCS method [12], which integrates a binary-addition-tree algorithm (BAT) [15] with stochastic sampling [16]. The BAT algorithm offers structural guidance through its hierarchical state space exploration, effectively identifying critical system configurations [15]. However, the original BAT-MCS method is limited by its reliance on fixed simulation numbers and a relatively simplistic approach to selecting core sub-vectors (super-vectors) that represent essential state configurations, thereby missing opportunities for further optimization [12].

To address these limitations, we propose cBAT-MCS, an advanced variant incorporating two key innovations: (1) a stratified layer-cut strategy for targeted super-vector selection, and (2) a novel reliability approximation formula to enhance computational efficiency based on the novel concept called normalization factor. By prioritizing critical state vectors and deprioritizing less influential ones, this framework intelligently redistributes computational resources, significantly reducing redundant calculations while preserving accuracy.

The significance of this contribution lies not only in the improved computational efficiency but also in the broader applicability of the method to real-world systems. The cBAT-MCS algorithm demonstrates particular strength in handling networks with heterogeneous component reliability characteristics and complex structural dependencies, situations where traditional methods often struggle to maintain accuracy while managing computational resources effectively.

Extensive numerical experiments on large-scale binary-state networks demonstrate that cBAT-MCS consistently outperforms both traditional MCS and the original BAT-MCS in terms of computational efficiency and accuracy. These results establish cBAT-MCS as a robust and scalable simulation tool for analyzing complex network systems, with potential applications across various domains in reliability engineering and system optimization.

The paper is structured as follows: Section 2 defines key acronyms, notations, nomenclature, and assumptions foundational to the study. Section 3 reviews related work and fundamental concepts in MCS, BAT, path-based layered search algorithm (PLSA), and BAT-MCS, providing essential background on current approaches and their limitations. Section 4 presents the cBAT-MCS methodology, detailing its layer-cut strategy, optimal super-cut selection, and reliability approximation formulation for computational efficiency optimization. Section 5 presents experimental benchmarks, performance metrics (e.g., accuracy, runtime), and sensitivity analyses against existing approaches. Section 6 concludes with key findings and future directions, such as multi-state system extensions and dynamic reliability frameworks.

## 2. ACRONYMS, NOTATIONS, NOMENCLATURE, AND ASSUMPTIONS

This section outlines the essential acronyms, notations and definitions, and assumations that form the foundation of the proposed cBAT-MCS methodology.

### 2.1 Acronyms

MCS: Monte Carlo simulation

BAT: Binary-Addition-Tree Algorithm

BAT-MCS: BAT-based MCS

cBAT-MCS: Proposed new cut-based BAT-MCS

PLSA: Path-based LSA

### 2.2 Mathematical Notations and Definitions

### 2.2.1 Basic Mathematical Operations

$|\bullet|$: Cardinality (number of elements in a set)

$E[\bullet]$: Expected value

$\text{Var}[\bullet]$: Variance

$\text{Pr}^*(\bullet)$: Approximate probability

$a_i$: arc $i$

$V$: $V = \{1, 2, \ldots, n\}$

$E$: $E = \{a_1, a_2, \ldots, a_m\}$

$G(V, E)$: an undirected graph with $V$ and $E$. For example, **Figure 1** is a graph with $V = \{1, 2, 3, 4\}$, $E = \{a_1, a_2, a_3, a_4, a_5\}$, source node 1, and sink node 4

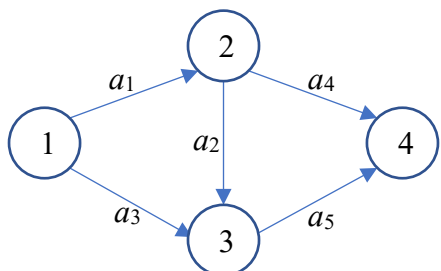

**Figure 1.** Example network

### 2.2.2 Network Components

$\Pr(a_i)$: probability of arc $a_i$ being in state 1 (working). For example, $\Pr(a_1) = 0.9$, $\Pr(a_2) = 0.8$, $\Pr(a_3) = 0.9$, $\Pr(a_3) = 0.95$, $\Pr(a_4) = 0.85$, $\Pr(a_5) = 0.8$ in **Figure 1**.

$X$: state vector/supervector

$\underline{X}$: the set or all supervectors where arc $a$ is simultaneously present ($a \in \underline{X}$) and absent ($a \notin X$) in supervector $X$.

$X(a_i)$: state of arc $a_i$ in vector $X$ for $i = 1, 2, \ldots, m$

$X(a_{[i]})$: the $i$th arc selected in vector/supervector $X$

$\beta$: number of arcs in the supervector for BAT-MCS

$b$: number of arcs in the layer-cut for the proposed cBAT-MCS

$\mathbf{D}$: state distribution $\mathbf{D}(a) = \Pr(a)$ for all $a \in E$, e.g., $\mathbf{D}$ is listed in **Table 1**.

**Table 1.** Probability distribution $\mathbf{D}$.

| $i$ | $\mathbf{D}(a_i) = \Pr(a_i)$ |
|---|---|
| 1 | 0.9 |
| 2 | 0.8 |
| 3 | 0.7 |
| 4 | 0.6 |
| 5 | 0.5 |

$G(V, E, \mathbf{D})$: a binary-state network can be modeled as a graph $G(V, E)$ with a state distribution $\mathbf{D}$. $G(V, E)$ with the $\mathbf{D}$. For example, **Figure 1** illustrates such a network configuration where $\mathbf{D}$, as specified in **Table 1**, governs the system's operational states.

$G(X)$: the subnetwork of $G$ is defined as $G(X) = G(V, E(X), \mathbf{D})$, where $E(X) = \{a \in E \mid X(a) = 1\}$.

$0(X)$: A zero subvector such the state of each arc in $X$ is in a failed state (i.e., $x_i = 0$ for all $x_i \in X$).

### 2.2.3 Simulation Representations and Parameters

$R_{MCS}$: Reliability estimate derived from Monte Carlo simulation (MCS).

$R_{BAT-MCS}$: Reliability estimate obtained via the proposed BAT-MCS method.

$R_{cBAT-MCS}$: Reliability estimate obtained via the proposed cBAT-MCS method.

$R(G)$: Exact reliability of the network $G(V, E, \mathbf{D})$, where $\mathbf{D}$ denotes the state distribution.

$\Omega(C)$: $\{X \mid$ for all supervector $X$ are solutions of arc set $C\}$

$\Omega_1(C)$: $\Omega_1(C) = \{X \mid X = (x_{[1]}, x_{[2]}, \ldots, x_{[b]}) \in \Omega(C), x_{[k]} \in \{0, 1\}, \sum_{k=1}^{b} X(a_{[k]}) \geq 1\}$

$N_{sim}$: Total number of simulations in a single run

$N_{sim}(X)$: Number of simulations for a specific supervector $X$ in a single run

$N_{pass}$: Total number of connected vectors identified across all simulations in a single run

$N_{pass}(X)$: Number of connected vectors identified for a specific supervector $X$ in a single run

$N_{run}$: Total number of total runs

## 2.3 Nomenclature

Binary-state network: A network in which every arc (or link) can exist in one of two states, typically represented as 0 (failed) or 1 (operational).

Reliability: The probability that a network operates successfully, meaning it provides a functioning path between specified nodes in a binary-state network.

Supervector: A supervector is defined as a sub-vector $X = (x_1, x_2, \ldots, x_\beta)$ such that for each index $i$ (where $i \leq \beta \leq m$), the relation $X(a_i) = x_i$ holds [36].

Connect/disconnect vector: A vector $X$ is a connected/ disconnect if nodes 1 and n is connected in $G(X)$.

Subnetwork $G(X)$ is a subnetwork induced by the binary vector $X$. This means $G(X)$ contains all vertices $V$ from the original graph $G$, but only includes edges a for which the corresponding element in $X$ equals 1, while maintaining the same state distribution **D**. **Figure 2**. $G(X)$, where $X = (1, 1, 0, 1, 0)$ and $G(V, E)$ is shown in **Figure 1.** depicts the network configuration $G(X)$, where $X = (1, 1, 0, 1, 0)$ represents the binary-state vector defining the operational status of network components. This configuration is derived from the base graph $G(V, E)$ illustrated in **Figure 1**, with $X$ determining the active/inactive states of edges or nodes.

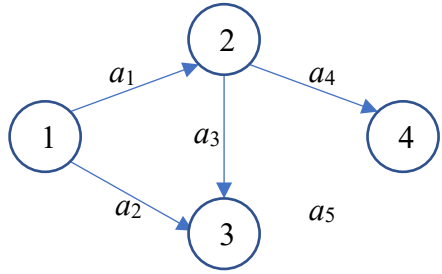

**Figure 2.** $G(X)$, where $X = (1, 1, 0, 1, 0)$ and $G(V, E)$ is shown in **Figure 1**.

## 2.4 Assumptions

1. Networks contain no parallel arcs or loops.

2. All nodes are assumed to be completely reliable and interconnected.

3. All arc states are statistically independent.

## 3. REVIEW OF BAT, PLSA, LAYER-CUTS, MCS, AND BAT-MCS

The proposed cBAT-MCS extends the BAT-MCS framework [12] by synthesizing four key methodologies: the BAT [15], PLSA [41], supervector [19], and MCS [21, 23, 34]. This section provides a systematic analysis of these foundational components and their integration, beginning with individual examinations of the theoretical foundations of BAT, core principles of PLSA, supervector, and statistical properties of MCS, followed by analysis of the unified BAT-MCS framework.

Special attention is given to the statistical advantages, specifically enhanced variance reduction and improved rare-event sampling capabilities, as well as the computational benefits of enhanced scalability and algorithmic efficiency achieved within the unified cBAT-MCS architecture.

### 3.1 BAT

The BAT, introduced by Yeh [15], provides a deterministic methodology for exhaustive state space enumeration in optimization problems, independent of solution feasibility. The algorithm initializes with a zero vector $X = (0, 0, ..., 0)$ and generates subsequent state vectors through iterative application of two fundamental rules [15]:

1. **Saturation Rule**: Locate the first failed arc (denoted as $a$) in $X$. Reset the states of all arcs preceding $a$ to zero and change the state of $a$ to working to form a new state vector.

2. **Termination Rule**: If no failed arc exists, the algorithm terminates.

The BAT algorithm's elegance derives from its fundamental design principles of simplicity, computational efficiency, and adaptability [15]. Through systematic state vector transitions, it achieves comprehensive exploration of configuration spaces while optimizing computational resources. The algorithm maintains $O(2^{m-1})$ time complexity [17] and minimal memory overhead by eliminating redundant state storage [18]. The complete algorithmic implementation is detailed in the pseudocode below, with the full source code available in [15].

The complete algorithmic implementation is detailed in the pseudocode below, with the full source code implementation available in [15].

**Algorithm: BAT**

**Input:** $m$ (the number of coordinates).

**Output:** All possible binary-state $m$-tuple vectors.

**STEP 0.** Initialize $X$ as a zero $m$-tuple vector.

**STEP 1.** Identify the first failed arc in $X$, say $a_k$. If no such arc exists, halts, and all $m$-tuple vectors have been generated.

**STEP 2.** Let $X(a_k) = 1$, reset $x_j = 0$ for $j = 1, 2, \ldots, k-1$, a new $X$ is generated, and go to STEP 1.

**Table 2** provides a detailed, step-by-step breakdown of the BAT procedure (as depicted in **Figure 1**), demonstrating its systematic exploration of binary-state vectors. The table includes columns $T(X)$, $C(X)$, $W(X)$, and $\text{Pr}_{0.9}(X_i)$, which represent illustrative examples of time, cost, weight, and probability functions associated with state vector $X$. These functions are included solely for demonstration purposes and do not represent actual implementation functions.

**Table 2.** All vectors obtained from the BAT.

| iteration | $X$ | $T(X)$ | $C(X)$ | $W(X)$ | $\text{Pr}_{0.9}(X_i)$ |
|---|---|---|---|---|---|
| 1 | (0, 0, 0, 0, 0) | 19 | 19 | 27 | 0.780255 |
| 2 | (1, 0, 0, 0, 0) | 12 | 13 | 29 | 0.845525 |
| 3 | (0, 1, 0, 0, 0) | 11 | 18 | 10 | 0.012021 |
| 4  | (1, 1, 0, 0, 0) | 14 | 21 | 24 | 0.775818 |
| 5  | (0, 0, 1, 0, 0) | 17 | 11 | 14 | 0.368082 |
| 6  | (1, 0, 1, 0, 0) | 13 | 10 | 16 | 0.576176 |
| 7  | (0, 1, 1, 0, 0) | 10 | 11 | 24 | 0.114396 |
| 8  | (1, 1, 1, 0, 0) | 18 | 28 | 38 | 0.189458 |
| 9  | (0, 0, 0, 1, 0) | 17 | 23 | 24 | 0.140470 |
| 10 | (1, 0, 0, 1, 0) | 16 | 16 | 10 | 0.49619 |
| 11 | (0, 1, 0, 1, 0) | 14 | 22 | 26 | 0.478777 |
| 12 | (1, 1, 0, 1, 0) | 19 | 22 | 25 | 0.604067 |
| 13 | (0, 0, 1, 1, 0) | 18 | 12 | 28 | 0.078764 |
| 14 | (1, 0, 1, 1, 0) | 11 | 11 | 25 | 0.719466 |
| 15 | (0, 1, 1, 1, 0) | 14 | 24 | 22 | 0.150774 |
| 16 | (1, 1, 1, 1, 0) | 12 | 11 | 15 | 0.824481 |
| 17 | (0, 0, 0, 0, 1) | 14 | 26 | 28 | 0.737104 |
| 18 | (1, 0, 0, 0, 1) | 17 | 16 | 36 | 0.340655 |
| 19 | (0, 1, 0, 0, 1) | 15 | 10 | 28 | 0.738322 |
| 20 | (1, 1, 0, 0, 1) | 11 | 26 | 15 | 0.659068 |
| 21 | (0, 0, 1, 0, 1) | 16 | 13 | 38 | 0.984819 |
| 22 | (1, 0, 1, 0, 1) | 19 | 16 | 31 | 0.333269 |
| 23 | (0, 1, 1, 0, 1) | 10 | 17 | 20 | 0.828836 |
| 24 | (1, 1, 1, 0, 1) | 18 | 29 | 27 | 0.704008 |
| 25 | (0, 0, 0, 1, 1) | 16 | 24 | 19 | 0.127426 |
| 26 | (1, 0, 0, 1, 1) | 14 | 13 | 28 | 0.635756 |
| 27 | (0, 1, 0, 1, 1) | 17 | 28 | 34 | 0.885850 |
| 28 | (1, 1, 0, 1, 1) | 12 | 29 | 28 | 0.717737 |
| 29 | (0, 0, 1, 1, 1) | 12 | 25 | 35 | 0.050024 |
| 30 | (1, 0, 1, 1, 1) | 19 | 22 | 16 | 0.658577 |
| 31 | (0, 1, 1, 1, 1) | 11 | 18 | 21 | 0.205589 |
| 32 | (1, 1, 1, 1, 1) | 15 | 28 | 31 | 0.863869 |

**3.2 PLSA in verifing the connectness and finding layers**

The Path-Based Layered Search Algorithm (PLSA) evolved from the Layered Search Algorithm (LSA) [19], which was originally designed for identifying *d*-minimal paths in acyclic networks [19]. While LSA's core framework offers computational efficiency and simplicity, it has spawned several specialized variants: PLSA for source-to-sink connectivity verification [15], Group Layered Search Algorithm (GLSA) for multi-node connectivity [1], and Total Layered Search Algorithm (TLSA) for full-network connectivity validation [20].

The PLSA, introduced in [15], adapts LSA's layered-search mechanism with a fundamental principle: any node n added to layers is necessarily connected to the source node (node 1). This approach eliminates redundant path evaluations, enhancing computational efficiency particularly for large-scale network reliability analyses [15]. The detailed implementation is presented in the pseudocode below [15].

**Algorithm: PLSA**

**Input:** A binary-state vector $X$, defining arc states in network $G(X)$.

**Output:** Determines whether nodes 1 and $n$ are connected in $G(X)$.

**STEP 0.** Initialize the first layer $L_1 = \{1\}$, $i = 2$, and $L_2 = \emptyset$.

**STEP 1.** Let layer $L_i = \{ v \in V \mid \text{if } X(e_{u,v}) > 0, u \in L_{i-1}, \text{ and } v \notin L_{i-1} \text{ in } G(X)\}$.

**STEP 2.** If $n \in L_i$, terminate and $G(X)$ is connected.

**STEP 3.** If $L_i = \emptyset$, terminate and $G(X)$ is disconnected. Otherwise, let $i = i + 1$, $L_i = \emptyset$, and go to STEP 1.

The PLSA implements a breadth-first search (BFS)-inspired traversal that partitions nodes into discrete layers while ensuring single-visit node exploration [41]. The algorithm achieves $O(n)$ time complexity, where $n$ represents the number of nodes, due to its guaranteed termination within $n-1$ iterations—the maximum possible path length in an $n$-node network. This linear complexity ensures efficient scalability for large-scale networks through systematic layer-wise expansion from the source node [41].

To demonstrate, consider the capacity vector $X = (2, 4, 4, 1, 4)$ from **Figure 1**. The PLSA's layer-wise expansion results, documented in the "$L_i$" column of **Table 3**, confirm network connectivity. **Table 4** extends this analysis to 16 vectors derived from two sets: four MCS-selected vectors $(x_3, x_4, x_5) \in \{(0, 1, 1), (1, 1, 0), (1, 1, 1), (1, 0, 1)\}$ (Section 3.3), and four supervector combinations $(x_1, x_3) \in \{(0,0), (1,0), (0,1), (1,1)\}$ for $(a_4, a_5)$ (Section 3.2).

Table 3. Finding all layers (via PLSA) and layer-cuts (discussed in Section 3.3).

| $i$ | $L_i$ | $c_i$ |
|---|---|---|
| 1 | $\{1\}$ | $\{a_1, a_2\}$ |
| 2 | $\{2, 3\}$ | $\{a_4, a_5\}$ |
| 3 | $\{4\}$ | |

Table 4. Verify the connectivity of 16 vectors using PLSA.

| $i$ | $x_1$ | $x_2$ | $x_3$ | $x_4$ | $x_5$ | Connect? |
|---|---|---|---|---|---|---|
| 1 | 0 | 1 | 1 | 0 | 0 | No |
| 2 | 1 | 1 | 0 | 0 | 0 | No |

| | | | | | | | |
|---|---|---|---|---|---|---|---|
| 3 | 1 | 1 | 1 | 0 | 0 | No |
| 4 | 1 | 0 | 1 | 0 | 0 | No |
| 5 | 0 | 1 | 1 | 1 | 0 | No |
| 6 | 1 | 1 | 0 | 1 | 0 | Yes |
| 7 | 1 | 1 | 1 | 1 | 0 | Yes |
| 8 | 1 | 0 | 1 | 1 | 0 | Yes |
| 9 | 0 | 1 | 1 | 0 | 1 | Yes |
| 10 | 1 | 1 | 0 | 0 | 1 | Yes |
| 11 | 1 | 1 | 1 | 0 | 1 | Yes |
| 12 | 1 | 0 | 1 | 0 | 1 | No |
| 13 | 0 | 1 | 1 | 1 | 1 | Yes |
| 14 | 1 | 1 | 0 | 1 | 1 | Yes |
| 15 | 1 | 1 | 1 | 1 | 1 | Yes |
| 16 | 1 | 0 | 1 | 1 | 1 | Yes |

**3.3 Supervector and its Superfamily**

The supervector $S_{super} = (a_1, a_2, \ldots, a_\beta)$ represents a specialized subset of vectors, where each position $i$ (ranging from 1 to $\beta \leq m$) denotes the operational state of the $i$-th arc [12, 19]. Originally introduced in the quick BAT algorithm [19] and later incorporated into BAT-MCS [12], this construct optimizes processing efficiency by facilitating rapid identification and exclusion of disconnected vectors during component connectivity verification.

The associated superfamily $\Omega(S_{super}) = \{X \mid X = (x_1, x_2, \ldots, x_\beta)$, where $x_i = 0, 1\}$ encompasses all possible solutions for the supervector $S_{super} = (a_1, a_2, \ldots, a_\beta)$. This concept, first proposed in BAT-MCS [12], enables computational optimization by efficiently partitioning the solution space into feasible ($F(X) = 1$) and infeasible ($F(X) = 0$) configurations.

For example, in **Figure 1**, where $S_{super} = (a_1, a_2)$ as a cut and subvector. The corresponding configuration space $\Omega(S_{super})$ comprises all possible combinations of $(x_4, x_5)$, defined as $\Omega(a_1, a_2) = \{ X \mid X = (x_1, x_2) = (0, 0), (1, 0), (0, 1), (1, 1)\}$, with cardinality $|\Omega(a_1, a_2)| = 2^2 = 4$.

**3.4 MCS and Its Important Statistical Characteristics**

Monte Carlo Simulation (MCS) is a widely adopted computational method for estimating solutions to high-dimensional, nonlinear, or computationally intractable problems [12, 21]. For this study, the crude MCS approach—as detailed in [22, 23]—was selected to approximate binary-state network reliability, chosen for its implementation simplicity and demonstrated effectiveness in reliability analysis [22, 23]. The algorithm estimates system behavior through statistical sampling of network states, achieving an optimal balance between computational efficiency and estimation accuracy [21].

Consider a binary-state stochastic network $G(V, E, \mathbf{D})$, where $V$ denotes the set of nodes, $E$ represents the directed edges, and $\mathbf{D}$ defines all arc reliability parameters, with $\mathbf{D}(a) = \Pr(a)$ indicating the probability that arc $a$ is operational. The network includes a source node $s = 1$ and a sink node $t = n$. Additionally, let $N_{sim}$ denote the predefined number of MCS to be performed. The pseudocode for this crude MCS implementation follows [24].

**Algorithm: MCS**

**Input:** $G(V, E, \mathbf{D})$ and $N_{sim}$.

**Output:** Approximated binary-state netork reliability $R_{MCS}$.

**STEP 0.** Let $sim = i = 1$ and $N_{pass} = 0$.

**STEP 1.** Let $X(a_i) = 1$ if $\rho \leq \Pr(a_i)$, else $X(a_i) = 0$, where $\rho$ is a uniformly distributed random number in $[0, 1]$.

**STEP 2.** If $i < m$, let $i = i + 1$ and go to STEP 1.

**STEP 3.** If $X$ is connected, let $N_{pass} = N_{pass} + 1$.

**STEP 4.** If $sim < N_{sim}$, let $sim = sim + 1$, $i = 1$, and go to STEP 1.

**STEP 5.** Let $R_{MCS} = N_{pass}/N_{sim}$ and halt.

The MCS estimator $R_{MCS}$ is unbiased and consistent [12], with

$$E[\sum_{j=1}^{N_{sim}} \frac{R_{MCS,j}}{N_{sim}}] = R \tag{1}$$

$$Var[R_{MCS}] = \sum_{j=1}^{N_{sim}} \frac{R(1-R)}{N_{sim}}, \tag{2}$$

where $R$ is the exact binary-state network reliability. To achieve a desired relative error $\epsilon$ at a $(1-\alpha)\%$ confidence level, the required number of simulations $N_{sim}$ satisfies [21, 24]:

$$\frac{z_{\frac{\alpha}{2}}^2(1-R)}{\epsilon^2 R} \leq N_{sim}, \tag{3}$$

where $z_{\alpha/2}$ is the critical value of the standard normal distribution. These properties are rigorously proven in foundational MCS literature [22, 23].

To illustrate the methodology, consider the network structure in **Figure 1**, where arc state probabilities are provided in **Table 1**. A MCS framework with $N_{sim} = 16$ trials was applied, and the results are summarized in **Table 5**. For each simulation, the value of $x_i$ (representing the state of arc $i$) is dynamically assigned based on the corresponding probability $\rho_i$, where $i = 1, 2, 3, 4, 5$, as outlined in **STEP 1** of the MCS pseudocode.

**Table 5** reveals a critical limitation arising from redundant sampling instances (e.g., $X_4 = X_{14}$, $X_8 = X_{13}$, $X_{11} = X_{16}$, $X_{12} = X_{15}$, $X_2 = X_5$, $X_7 = X_9$, where $X_i$ is the vector obtained in simulation trials $i$), which occur due to an insufficiently small $N_{sim}$. This inadequate sampling diversity results in the overrepresentation of specific system states, thereby compromising the accuracy of reliability estimates. These findings emphasize the imperative of selecting $N_{sim}$ in accordance with Eq. (3) to ensure statistical robustness.

The analysis reveals that $N_{pass} = 13$ because there are 13 simulated state vectors connected across all permutations of $X$. Consequently, $R_{MCS} = N_{pass}/N_{sim} = 12/16 = 0.750$ (see **Table 5**). This value diverges from the exact system reliability of 0.540, a discrepancy attributable to the limited simulation size ($N_{sim} = 16$) and the adoption of a simplistic MCS framework that fails to incorporate solution quality metrics.

Table 5. MCS Procedure for **Figure 1** based on **Table 1**.

| $i$ | $\rho_1$ | $\rho_2$ | $\rho_3$ | $\rho_4$ | $\rho_5$ | $x_1$ | $x_2$ | $x_3$ | $x_4$ | $x_5$ | Connect? |
|---|---|---|---|---|---|---|---|---|---|---|---|
| 1 | 0.92168 | 0.71974 | 0.73371 | 0.40070 | 0.15396 | 0 | 1 | 0 | 1 | 1 | No |
| 2 | 0.40533 | 0.33365 | 0.66730 | 0.83186 | 0.01404 | 1 | 1 | 1 | 0 | 1 | Yes |
| 3 | 0.71173 | 0.83719 | 0.66681 | 0.31019 | 0.47161 | 1 | 0 | 1 | 1 | 1 | Yes |
| 4 | 0.33772 | 0.36953 | 0.55257 | 0.62505 | 0.83836 | 1 | 1 | 1 | 0 | 0 | No |
| 5  | 0.39771 | 0.73576 | 0.35345 | 0.86882 | 0.08958 | 1 | 1 | 1 | 0 | 1 | Yes |
| 6  | 0.78790 | 0.43831 | 0.72763 | 0.14010 | 0.01628 | 1 | 1 | 0 | 1 | 1 | Yes |
| 7  | 0.56750 | 0.02068 | 0.04221 | 0.40355 | 0.28200 | 1 | 1 | 1 | 1 | 1 | Yes |
| 8  | 0.78631 | 0.79431 | 0.96743 | 0.57409 | 0.86514 | 1 | 1 | 0 | 1 | 0 | Yes |
| 9  | 0.34288 | 0.56388 | 0.52413 | 0.17044 | 0.41365 | 1 | 1 | 1 | 1 | 1 | Yes |
| 10 | 0.94191 | 0.66559 | 0.92689 | 0.86024 | 0.49946 | 0 | 1 | 0 | 0 | 1 | No  |
| 11 | 0.19682 | 0.65801 | 0.28775 | 0.15021 | 0.93827 | 1 | 1 | 1 | 1 | 0 | Yes |
| 12 | 0.31759 | 0.88889 | 0.38306 | 0.99749 | 0.31031 | 1 | 0 | 1 | 0 | 1 | Yes |
| 13 | 0.75611 | 0.33225 | 0.90883 | 0.33284 | 0.89182 | 1 | 1 | 0 | 1 | 0 | Yes |
| 14 | 0.02275 | 0.35458 | 0.20985 | 0.98929 | 0.54591 | 1 | 1 | 1 | 0 | 0 | No  |
| 15 | 0.20804 | 0.99488 | 0.07421 | 0.93896 | 0.18774 | 1 | 0 | 1 | 0 | 1 | Yes |
| 16 | 0.75760 | 0.00952 | 0.14548 | 0.54478 | 0.78840 | 1 | 1 | 1 | 1 | 0 | Yes |

### 3.5 BAT-MCS

In the traditional BAT-MCS framework [12], the system reliability $R_{\text{BAT-MCS}}$ is approximated using the following equation:

$$R_{\text{BAT-MCS}} = \sum_{\forall X} \Pr(X) \times \Pr^*(\underline{X}), \qquad (4)$$

where

- each supervector $X = (x_1, x_2, \ldots, x_\beta)$ is obtained from the BAT and its probaility is defined:

$$\Pr(X) = \sum_{\forall x_k=1} \Pr(a_k) \times \sum_{\forall x_k=0} [1 - \Pr(a_k)], \qquad (5)$$

- $\Pr_{\text{MCS}}(\underline{X})$ is the approximated probaility obtained from the MCS after known $X$:

$$\Pr_{\text{MCS}}(\underline{X}) = \frac{N_{\text{pass}}(X)}{N_{\text{sim}}(X)}, \qquad (6)$$

- the total number of simulations allocated to $X$:

$$N_{\text{sim}}(X) = N_{\text{sim}} \times \frac{\Pr(X)}{\sum_{\forall X} \Pr(X)}. \qquad (7)$$

Eq. (4) aggregates the contributions of all vectors $X$, weighted by their respective probabilities and simulation-based failure rates. Computing $\Pr(X)$ using Eq. (5) requires $b$ multiplications to obtain $\Pr(X)$ for each possible $X$. For $\Pr_{\text{MCS}}(X)$ in Eq. (6), each calculation requires one division operation. Calculating $N_{\text{sim}}(X)$ involves one multiplication operation, followed by another multiplication after obtaining $\sum_{\forall X} \Pr(X)$. Therefore, the total computational requirements for $R_{\text{BAT-MCS}}$ in Eq. (4) sum to:

- Multiplications: $(\beta \times 2^\beta + 2 \times 2^\beta + 1)$ operations
- Divisions: $2^\beta + 1$ operations
- Summations: $2^\beta + 1$ operations

A key limitation of this approach is the variance of $R_{\text{BAT-MCS}}$ [12]:

$$Var[R_{\text{BAT-MCS}}] = \frac{R - 2^\beta R^2}{\lambda}, \text{ where } \lambda = N_{\text{sim}}/2^\beta \qquad (8)$$

cannot be analytically derived from the equation. This necessitates the use of empirical methods for error estimation.

To elucidate the fundamental distinctions between conventional MCS and the BAT-MCS, we revisit the network in **Figure 1** under identical experimental conditions:

- The simulation size is fixed at $N_{\text{sim}} = 16$ trials for direct methodological comparison.
- Arc state probabilities $\Pr(a_i)$ for $i=1, 2, 3, 4, 5$ are retained from Table 1
- Uniformly distributed random number $\rho_i \in [0, 1]$ for $i=1, 2, 3, 4, 5$ are from **STEP 5**.

In conventional MCS, each trial independently samples the state $x_i$ of all arcs ($i = 1, 2, 3, 4, 5$) using their respective $\rho_i$. In contrast, BAT-MCS strategically partitions the problem into supervector and its family, e.g., supervector $S_{\text{super}} = (a_1, a_2)$ and superfamily $\Omega(S_{\text{super}}) = \{(x_1, x_2) \mid (0, 0), (1, 0), (0, 1), (1, 1)\}$).

As demonstrated in **Table 6**, BAT-MCS evaluates these three supervectors $(x_1, x_2) = (0, 0)$, $(1, 0)$, $(0, 1)$, and $(1, 1)$ alongside their joint probabilities $\Pr(X)$ and allocated simulation counts $N_{\text{sim}}(X)$.

Table 6. $N_{\text{sim}}(X)$ and $\Pr(X)$ for **Figure 1** based on **Table 1**.

| $X = (x_1, x_2)$ | (0, 0) | (1, 0) | (0, 1) | (1, 1) |
|---|---|---|---|---|
| $\Pr(X)$ | 0.1×0.2=0.02 | 0.9×0.2=0.18 | 0.1×0.8=0.08 | 0.9×0.8=0.72 |
| $\Pr(X)/\sum_X \Pr(X)$ | 0.02 | 0.18 | 0.08 | 0.72 |
| $N_{\text{sim}} \times \Pr(X)/\sum_X \Pr(X)$ | 0.32 | 2.88 | 1.28 | 11.52 |
| $N_{\text{sim}}(X)$ | 1 | 3 | 2 | 10 |

Table 7. BAT-MCS for **Figure 1** based on **Table 5** and **Table 6**.

| $(x_1, x_2)$ | $i$ | $\rho_3$ | $\rho_4$ | $\rho_5$ | $x_3$ | $x_4$ | $x_5$ | Connect? | $N_{pass}(X)/N_{sim}(X)$ |
|---|---|---|---|---|---|---|---|---|---|
| (0, 0) | 1 | 0.73371 | 0.40070 | 0.15396 | 0 | 1 | 1 | No | 0/1 |
| (1, 0) | 1 | 0.66730 | 0.83186 | 0.01404 | 1 | 0 | 1 | Yes | 2/3 |
|  | 2 | 0.66681 | 0.31019 | 0.47161 | 1 | 1 | 1 | Yes | |
|  | 3 | 0.55257 | 0.62505 | 0.83836 | 1 | 0 | 0 | No | |
| (0, 1) | 1 | 0.35345 | 0.86882 | 0.08958 | 1 | 0 | 1 | Yes | 1/2 |
|  | 2 | 0.72763 | 0.14010 | 0.01628 | 0 | 1 | 1 | No | |
| (1, 1) | 1 | 0.04221 | 0.40355 | 0.28200 | 1 | 1 | 1 | Yes | 8/10 |
|  | 2 | 0.96743 | 0.57409 | 0.86514 | 0 | 1 | 0 | Yes | |
|  | 3 | 0.52413 | 0.17044 | 0.41365 | 1 | 1 | 1 | Yes | |
|  | 4 | 0.92689 | 0.86024 | 0.49946 | 0 | 0 | 1 | No | |
|  | 5 | 0.28775 | 0.15021 | 0.93827 | 1 | 1 | 0 | Yes | |
|  | 6 | 0.38306 | 0.99749 | 0.31031 | 1 | 0 | 1 | Yes | |
|  | 7 | 0.90883 | 0.33284 | 0.89182 | 0 | 1 | 0 | Yes | |
|  | 8 | 0.20985 | 0.98929 | 0.54591 | 1 | 0 | 0 | No | |
|  | 9 | 0.07421 | 0.93896 | 0.18774 | 1 | 0 | 1 | Yes | |
|  | 10 | 0.14548 | 0.54478 | 0.78840 | 1 | 1 | 0 | Yes | |

Applying Eq. (4) in BAT-MCS based on the results from **Table 7** yields:

$$R_{\text{BAT-MCS}} = 0.02 \times 0/1 + 0.18 \times 2/3 + 0.08 \times 1/2 + 0.72 \times 8/10 = 0.7360. \tag{9}$$

Despite the limited simulation size ($N_{sim} = 16$), the $R_{\text{BAT-MCS}}$ result (0.7360) demonstrates closer alignment with the exact system reliability (0.72461) compared to the conventional MCS estimate ($R_{\text{MCS}} = 0.8125$, **Table 5**). This improvement arises from BAT-MCS's variance-reduction strategy, which strategically prioritizes high-probability supervectors (e.g., $(x_1, x_2) = (1, 1) = 0.72$) while excluding deterministic failures (e.g., network cuts like $(x_1, x_2) = (0, 0)$).

## 4. THE PROPOSED cBAT-MCS

The cBAT-MCS framework is introduced as a computational engine to integrate deterministic and stochastic methods. The deterministic component leverages layer-cut decomposition derived from PLSA and its associated family of cuts generated via BAT. The stochastic component employs an MCS scheme, augmented with a novel approximation formula to compute system reliability efficiently. This hybrid methodology synthesizes structural insights from PLSA/BAT with probabilistic sampling via MCS, ensuring both computational tractability and statistical fidelity. The interplay of these components, along with their theoretical and algorithmic implications, is rigorously discussed in this section.

**4.1 Layer-Cut**

For an undirected graph $G(V, E)$, a cut is a partition of the node set $V$ into two disjoint node subsets $S$ and $T$ such that $S \cup T = V$ and $S \cap T = \emptyset$. A layer-cut is a specialized cut derived from the layered structure generated by the PLSA [15]. Specifically, it consists of arcs connecting nodes in two consecutive layers $L_i$ and $L_{i+1}$ obtained during PLSA's traversal. This concept, first proposed in [15], leverages the PLSA's iterative layer expansion to systematically identify critical edge sets that partition the graph. The pseudocode in finding all layer-cuts is listed below:

**Algorithm:** Finding All Layer-Cuts

**Input:**    A graph $G(V, E)$, the source node 1, and the sink node $n$.

**Output:**   All layer-cuts.

**STEP 0.**   Find all layers $L_1, L_2, \ldots, L_\lambda$ using PLSA.

**STEP 1.**   Find all layer-cuts $c_k = \{e_{i,j} \mid \text{nodes } i \text{ and } j \text{ are in } L_k \text{ and } L_{k+1}, \text{seperatedly}\}$ for $k = 1, 2, \ldots, (\lambda-1)$.

**STEP 2.**   Let $c_\lambda = E - \bigcup_{k=1}^{\lambda-1} c_k$.

The computational complexity for identifying all layer-cuts is $O(n+m)$, where $n$ is the number of nodes and $m$ is the number of arcs. Specifically, STEP 0 implements the PLSA algorithm with $O(n)$ complexity, as detailed in Section 3.2, while STEPs 1 and 2 require $O(m)$ operations as they traverse each arc exactly once.

To illustrate this process, consider the network depicted in **Figure 1**. The PLSA algorithm identifies three distinct layers: $L_1 = \{1\}$, $L_2 = \{2, 3\}$, and $L_3 = \{4\}$, as shown in the "$L_i$" column of **Table 3**. This layering structure yields two layer-cuts: $c_1 = \{a_1, a_2\}$ and $c_2 = \{a_4, a_5\}$, presented in **Table 3**'s rightmost column. Both $c_1$ and $c_2$ represent valid cuts in **Figure 1**, as they effectively partition the graph into disjoint node subsets.

## 4.2 Cut Superfamily

The theoretical bridge between deterministic and stochastic approaches in the proposed cBAT-MCS is established through cut-based superfamilies. On the contrast, the conventional BAT-MCS framework employs an oversimplified selection criterion for supervectors, which may compromise computational efficiency and solution accuracy. To address this limitation, we propose cBAT-MCS, a novel method grounded in the layer-cut superfamily—a concept introduced herein to systematically identify and discard infeasible superfamilies, thereby enhancing algorithmic performance.

The layer-cut superfamily synthesizes three methodologies: PLSA (Section 3.2), layer-cut concept (Section 5.1), and superfamily generation (Section 3.3). Formally, for a given layer-cut $C_{super} = (a_{[1]}, a_{[2]}, \ldots, a_{[b]})$ —a set of arcs whose removal disconnects source node 1 and terminal node $n$—the associated layer-cut superfamily $\Omega_1(C_{super})$ is defined as:

$$\Omega_1(C_{\text{super}}) = \{X \mid X = (x_{[1]}, x_{[2]}, \ldots, x_{[b]}), x_{[k]} \in \{0,1\}, \sum_{k=1}^{b} X(a_{[k]}) \geq 1\}. \quad (10)$$

where $X(a_{[k]}) = x_{[k]}$ denotes the state of arc the $k$th arc $a_{[k]}$ in $X$. By construction, $\Omega_1(C_{super})$ enforces at least one active arc per layer-cut, reducing the solution space cardinality from $2^b$ (where $b$ is the total arc count) to a smaller set.

**Example:** Consider layer-cuts $(a_1, a_2)$ and $(a_4, a_5)$ from Section 5.1. The cut superfamily for both cases is $\Omega_1(a_1, a_2) = \Omega_1(a_4, a_5) = \{(0, 1), (1, 0), (1, 1)\}$, which contains only 3 elements compared to the $2^2 = 4$ possible combinations. This demonstrates the inherent reduction capability of the cut superfamily, pruning infeasible configurations a priori.

This methodological advancement ensures rigorous feasibility guarantees while streamlining computational complexity—a critical advantage for large-scale network analyses.

## 4.3 The Normalization Factor

While cBAT-MCS retains the core reliability approximation framework of BAT-MCS (Eq. (4)), it introduces critical modifications to the vector definition and sampling strategy to enhance computational efficiency.

In BAT-MCS, the system state $X$ is represented as a subvector $X = (x_1, x_2, \ldots, x_\beta)$, where $\beta$ denotes the number of arcs in a single supervector. By contrast, cBAT-MCS redefines $X$ as a full-dimensional layer-cut $X = (x_{[1]}, x_{[2]}, \ldots, x_{[b]})$, where $b$ corresponds to the total number of arcs in the layer-cut topology, constrained by the layer-cut superfamily $\Omega_1(C_{\text{super}})$. This ensures $X \in \Omega_1(C_{\text{super}})$, where $C_{\text{super}} = (a_{[1]}, a_{[2]}, \ldots, a_{[b]})$ represents a layer-cut derived from PLSA, and guarantees the exclusion of trivial/infeasible states (e.g., the zero vector $(0, 0, \ldots, 0)$), as

$$\sum_{k=1}^{b} x_{[k]} \geq 1, \tag{11}$$

is enforced by construction.

To optimize computational efficiency and quantify the statistical variance of BAT-MCS, we propose a novel mathematical formulation under the assumption that

$$\gamma = \frac{\sum_{X \in \Omega_1(C_{\text{super}})} \Pr(X)}{N_{\text{sim}}} = \frac{1 - \Pr(0(C_{\text{super}}))}{N_{\text{sim}}} \tag{12}$$

yields an integer value. To optimize computational efficiency and quantify the statistical variance of the results, we propose a novel mathematical formulation:

$$R_{\text{cBAT-MCS}} = \sum_{X \in \Omega_1(C_{\text{super}})} [\Pr(X) \times \Pr_{\text{MCS}}(\underline{X})]$$

$$= \sum_{X \in \Omega_1(C_{\text{super}})} \left[ \Pr(X) \times \frac{N_{\text{pass}}(X)}{N_{\text{sim}}(X)} \right]$$

$$= \sum_{X \in \Omega_1(C_{\text{super}})} \left[ \Pr(X) \times \frac{N_{\text{pass}}(X)}{N_{\text{sim}} \times \frac{\Pr(X)}{\sum_{X \in \Omega_1(C_{\text{super}})} \Pr(X)}} \right]$$

$$= \sum_{X \in \Omega_1(C_{\text{super}})} \left[ \sum_{X \in \Omega_1(C_{\text{super}})} \Pr(X) \times \frac{N_{\text{pass}}(X)}{N_{\text{sim}}} \right]$$

$$= \sum_{X \in \Omega_1(C_{\text{super}})} \Pr(X) \times \sum_{X \in \Omega_1(C_{\text{super}})} \frac{N_{\text{pass}}(X)}{N_{\text{sim}}}$$

$$= \gamma \left[ \sum_{X \in \Omega_1(C_{\text{super}})} N_{\text{pass}}(X) \right]$$

$$= \gamma N_{\text{pass}}. \tag{13}$$

The computational efficiency is achieved through a single calculation of $\frac{\sum_{X \in \Omega} \Pr(X)}{N_{\text{sim}}}$ requiring $|\Omega| = 2^b - 1$ summations and one division operation. The total computational cost comprises one multiplication, one division, and $2|\Omega|$ summations. This demonstrates that calculating $R_{\text{cBAT-MCS}}$ is computationally more efficient than computing $R_{\text{BAT-MCS}}$, which takes $(\beta \times 2^\beta + 2 \times 2^\beta + 1)$ multiplications operations, $2^\beta + 1$ divisions operations, and $2^\beta + 1$ summations operations.

Furthermore, Eq. (12) facilitates the direct calculation of variance for the reliability approximation obtained through cBAT-MAS:

$$\text{Var}[R_{\text{cBAT-MCS}}] = \text{Var}[\sum_{X \in \Omega_1(C_{\text{super}})} \Pr(X) \times \sum_{X \in \Omega_1(C_{\text{super}})} \frac{N_{\text{pass}}(X)}{N_{\text{sim}}}]$$

$$= \left[ \sum_{X \in \Omega_1(C_{\text{super}})} \Pr(X) \right]^2 \text{Var}[\sum_{X \in \Omega_1(C_{\text{super}})} \frac{N_{\text{pass}}(X)}{N_{\text{sim}}}]$$

$$= \left[ \sum_{X \in \Omega_1(C_{\text{super}})} \Pr(X) \right]^2 \text{Var}[R]. \tag{14}$$

where $\Pr(X)$ is the reliability of state $X$. This formulation leverages the reduced cardinality of $\Omega(C_{\text{super}})$ (achieved by discarding infeasible states) to minimize redundant computations while preserving statistical rigor.

### 4.4 Determine Maximal-Probability Minimal-Size $C_{\text{super}}$ based on the Eq. (13)

In the proposed cBAT-MCS framework, the selection of the critical superset cut $C_{\text{super}}$ from available layer-cuts is pivotal to enhancing computational efficiency. Let $C_{\text{super}} = (a_{[1]}, a_{[2]}, \ldots, a_{[b]})$ represent a layer-cut identified via PLSA. As derived from Eq. (13), the probability $\Pr(0(C_{\text{super}})) = 1 - \sum_{X \in \Omega_1(C_{\text{super}})} \Pr(X) = \prod_{k=1}^{b}[1 - \Pr(a_{[k]})]$ governs the variance reduction mechanism: an increase in $\Pr(0(C_{\text{super}}))$ directly reduces both $\sum_{X \in \Omega_1(C_{\text{super}})} \Pr(X)$ and the variance

Var[$R_{cBAT-MCS}$].

The failure state set $\Omega_1(C_{super})$ has a cardinality bounded by $2^{b-1}$, where $b$ denotes the number of arcs in $C_{super}$. This bound establishes a direct exponential relationship between $b$ and the size of $\Omega_1(C_{super})$: as $b$ increases, the number of $X$ in $\Omega_1(C_{super})$ grows exponentially. While this introduces computational challenges, the expanded set enables a more comprehensive exploration of $X$. Specifically, for a fixed total number of MCS trials $N_{sim}$, the larger $\Omega_1(C_{super})$ allows finer-grained sampling across $\underline{X}$, improving the statistical resolution of the reliability estimate $R_{cBAT-MCS}$. This enhanced coverage reduces variance in the estimator, particularly when $N_{sim}$ scales proportionally to $|\Omega_1(C_{super})|$, thereby balancing computational effort with accuracy.

From the above, the optimal layer-cut selection follows a two-tier hierarchical criteria system:

- Primary Criterion: Minimize the number of arcs in $C_{super}$, i.e., $|C_{super}|$.
- Secondary Criterion: Among layer-cuts with equal number of arcs, select the one with maximal $\Pr(0(C_{super}))$, where each coordinate in $\Pr(0(C_{super}))$ represents the probability of inactive states.

**Algorithm:** Finding the Maximal-Probability Minimal-Size Layer-Cut

**Input:** A graph $G(V, E)$, all layer-cuts $c_1, c_2, \ldots, c_{\lambda-1}$, the source node 1, and the sink node $n$.

**Output:** The maximal layer-cut $C_{super}$.

**STEP 0.** Find $c^*$ such that $\|c\| \leq \|c^*\|$ for all $c^*$ and $c \in \{c_1, c_2, \ldots, c_{\lambda-1}\}$, where $\|c\|$ is the size metric (arcs) of a layer-cut $\|c\|$.

**STEP 1.** If there is only one $c^*$, let $C_{super} = c^*$ and halt.

**STEP 2.** Find the layer-cut $C_{super} = c_b$ with the maximal probability among these $c^*$ such that $\|c_b\| = \|c^*\|$ for all $c_b$ and $c^*$.

This hierarchical approach optimizes both variance reduction and computational efficiency. To illustrate this selection process, we examine two candidate layer-cuts from Section 3.3: $c_1 = \{a_1, a_2\}$ and $c_2 = \{a_4, a_5\}$. Computing the joint inactive probabilities using the binary-state probabilities presented in Table 1:

- For $c_1$: $\Pr(x_1=0, x_2=0)=(1−0.9)×(1−0.8)=0.1×0.2=0.02$.
- For $c_2$: $\Pr(x_4=0, x_5=0)=(1−0.6)×(1−0.5)=0.4×0.5=0.20$.

While $c_2$ exhibits a higer joint probability (0.20), indicating a potentially more effective variance reduction target, $c_2$ is ultimately selected as $C_{\text{super}}$. This selection reflects the hierarchical framework's emphasis on balancing computational tractability with probability-based optimization criteria.

**4.5 Pseudocode for cBAT-MCS Reliability Approximation**

The pseudocode for computing the approximate reliability $R_{\text{cBAT-MCS}}$ of a binary-state network using the cBAT-MCS framework is presented below. This algorithm integrates Eqs. (12) and (13), combining deterministic layer-cut superfamily analysis (Section 4.4) with stochastic Monte Carlo sampling (Section 4.3) to balance computational efficiency and statistical rigor.

**Algorithm: cBAT-MCS**

**Input:** A binary-state stochastic network $G(V, E, \mathbf{D})$, where:

$V$: Set of nodes, with source node $s = 1$ and sink node $t = n$.

$E$: Set of directed arcs.

$\mathbf{D}(a) = \Pr(a)$ for all adenotes the operational probability of arc $a$ for all $a \in E$.

$N_{\text{sim}}$: Predefined number of Monte Carlo simulations.

**Output:** Approximate reliability $R_{\text{cBAT-MCS}}$.

**STEP 0.** Identification of the minimal layer-cut and initial Probability Estimation:

- Identify the minimal layer-cut $C_{\text{super}} = (a_{[1]}, a_{[2]}, \ldots, a_{[b]})$ using the algorithm in Section 4.4.
- Determine the superfamily $\Omega_1(C_{\text{super}})$ defined in Eq. (10) using BAT.
- For each $X \in \Omega_1(C_{\text{super}})$, calculate the probability of supervector $\Pr(X)$.
- Calculate the normalization factor $\gamma = \dfrac{\sum_{X \in \Omega_1(C_{\text{super}})} \Pr(X)}{N_{\text{sim}}}$.

**STEP 1.** Aggregate results: $N_{pass} = \sum_{X \in \Omega_1(C_{super})} N_{pass}(X)$ using the MCS discussed in Section 3.4.

**STEP 2.** Compute the approximate reliability $R_{cBAT-MCS} = \gamma N_{pass}$ and halt.

The cBAT-MCS algorithm exhibits a total time complexity of $O(nN_{sim}(m-b)2^{b-1})$, dominated by STEP 1. In STEP 0, identifying the maximal layer-cut $C_{super}$ (size $b$), computing its superfamily $\Omega_1(C_{super})$, (cardinality $2^{b-1}$), and calculating the normalization factor $\gamma$ contribute $O(b2^{b-1})$, primarily due to the enumeration and probabilistic evaluation of $2^{b-1}$ supervectors in $\Omega_1(C_{super})$.

STEP 1, however, incurs $O(nN_{sim}(m-b)2^{b-1})$ operations, as it involves $N_{sim}$ MCS trials for each supervector $X \in \Omega_1(C_{super})$, requiring $(m-b)$ random variable generations and $O(n)$-time connectivity verification via the PLSA per trial. STEP 2, which computes the final reliability $R_{cBAT-MCS}$ through a scalar multiplication, adds negligible $O(1)$ overhead.

The exponential dependence on $b$ and linear scaling with $n$, $m$, and $N_{sim}$ highlight the method's tractability for systems with the maximal-probability minimal-size $C_{super}$ and its suitability for large-scale networks when $b$ is bounded. The size of $C_{super}$ is a critical factor in the algorithm's time complexity. The selection of $b = |C_{super}|$ significantly impacts both computational efficiency and estimator variance.

While the proposed cBAT-MCS and BAT-MCS share similar asymptotic time complexity, cBAT-MCS demonstrates superior operational efficiency. Specifically, cBAT-MCS reduces the number of multiplicative and summative operations by $2^b$ compared to BAT-MCS, as detailed in Section 4.3. Fewer operations minimize cumulative numerical errors, enhancing precision. In terms of variance, cBAT-MCS is more robust due to the inequality:

$$\left[\sum_{X \in \Omega_1(C_{super})} \Pr(X)\right]^2 \leq \left[\sum_{X \in \Omega(S_{super})} \Pr(X)\right]^2. \tag{15}$$

which reflects tighter probability bounds on the critical failure modes.

## 4.6 cBAT-MCS in Step-By-Step Procedure

To systematically investigate the fundamental differences among conventional MCS, BAT-MCS, and the proposed cBAT-MCS approaches, we conducted a comparative analysis using the network depicted in **Figure 1** under controlled experimental conditions. We standardized all methods by maintaining a consistent set of parameters. Our experimental framework involved 16 trials ($N_{sim}$ = 16) to allow for direct methodological comparisons. The arc state probabilities Pr($a$) for all arcs $a$ were preserved as specified in Table 1, and uniformly distributed random numbers $\rho_i$ in the interval [0, 1] for $i$ in {1, 2, 3, 4, 5} were generated according to **Table 5** of the protocol.

cBAT-MCS selectes the maximal-probability minimal-size $C_{super}$ = {$a_4$, $a_5$} from Sectin 4.4, and evaluates three supervectors in $\Omega_1(C_{super})$ = {($x_4$, $x_5$) = (1, 0), (0, 1), (1, 1)}, with their respective probabilities Pr($X$): 0.4 × 0.5 = 0.2, 0.6 × 0.5 = 0.3, 0.4 × 0.5 = 0.2, and 0.6 × 0.5 = 0.3, respectively. Applying Eq. (14), we have γ in cBAT-MCS yields:

$$\gamma = (0.30 + 0.2 + 0.3) / 16 = 0.05. \tag{16}$$

The application of Eq. (13) to the proposed cBAT-MCS framework, utilizing the computational results presented in **Table 10**, yields the following quantitative expression:

$$R_{cBAT\text{-}MCS} = \gamma N_{pass} = 0.05 \times (4 + 3 + 6) = 0.650. \tag{17}$$

The proposed cBAT-MCS framework demonstrates superior accuracy compared to conventional MCS and BAT-MCS methods, as evidenced by the following reliability indices computed for a benchmark system with exact reliability $R$ = 0.540:

$$|R - R_{cBAT\text{-}MCS}| = 0.11 < |R - R_{BAT\text{-}MCS}| = 0.196 < |R_{BAT} - R_{MCS}| = 0.21.$$

The proposed cBAT-MCS reduces absolute error by (0.196−0.11)/0.196 = 43.8% compared to BAT-MCS and 47.6% compared to conventional MCS. This improvement stems from its targeted sampling strategy and variance reduction mechanism, which refine the reliability estimation process by focusing on $\Omega_1(C_{super})$.

Table 8. cBAT-MCS for **Figure 1**.

| ($x_4$, $x_5$) | $i$ | $\rho_1$ | $\rho_2$ | $\rho_3$ | $x_1$ | $x_2$ | $x_3$ | Connect? | $N_{pass}(X)$ |
|---|---|---|---|---|---|---|---|---|---|
| (1, 0) | 1 | 0.92168 | 0.71974 | 0.73371 | 0 | 1 | 0 | No | 4 |

|       |   |         |         |         |   |   |   | Yes |   |
|-------|---|---------|---------|---------|---|---|---|-----|---|
|       | 2 | 0.40533 | 0.33365 | 0.66730 | 1 | 1 | 1 | Yes |   |
|       | 3 | 0.71173 | 0.83719 | 0.66681 | 1 | 0 | 1 | Yes |   |
|       | 4 | 0.33772 | 0.36953 | 0.55257 | 1 | 1 | 1 | Yes |   |
|       | 5 | 0.39771 | 0.73576 | 0.35345 | 1 | 1 | 1 | Yes |   |
|       | 6 | 0.78790 | 0.43831 | 0.72763 | 1 | 1 | 0 | Yes |   |
| (0, 1)| 1 | 0.56750 | 0.02068 | 0.04221 | 1 | 1 | 1 | Yes | 3 |
|       | 2 | 0.78631 | 0.79431 | 0.96743 | 1 | 1 | 0 | No  |   |
|       | 3 | 0.34288 | 0.56388 | 0.52413 | 1 | 1 | 1 | Yes |   |
|       | 4 | 0.94191 | 0.66559 | 0.92689 | 0 | 1 | 0 | No  |   |
| (1, 1)| 1 | 0.19682 | 0.65801 | 0.28775 | 1 | 1 | 1 | Yes | 6 |
|       | 2 | 0.31759 | 0.88889 | 0.38306 | 1 | 0 | 1 | Yes |   |
|       | 3 | 0.75611 | 0.33225 | 0.90883 | 1 | 1 | 0 | Yes |   |
|       | 4 | 0.02275 | 0.35458 | 0.20985 | 1 | 1 | 1 | Yes |   |
|       | 5 | 0.20804 | 0.99488 | 0.07421 | 1 | 0 | 1 | Yes |   |
|       | 6 | 0.75760 | 0.00952 | 0.14548 | 1 | 1 | 1 | Yes |   |

## 5. Method Performance Assessment: Boxplots, p-Values, and Absolute Errors

This section evaluates the performance of three reliability estimation methods—BAT-MCS, MCS, and cBAT-MCS—across three experimental configurations (Ex1, Ex2, Ex3) applied to 10 benchmark networks (IDs 1–10) as in **Figure 3**. Each method pair comparison is evaluated using two performance metrics: reliability (accuracy of estimates) and runtime (computational efficiency).

### 5.1 Experimental Framework

The computational experiments assessing the performance of cBAT-MCS, BAT-MCS, and MCS were conducted on an Intel Core i7-10750H CPU (2.60 GHz base frequency, 6 cores, 12 threads, 12 MB cache) with 64 GB DDR4 RAM, running a 64-bit Windows 11 Pro operating system. This high-performance setup ensured robust handling of large-scale simulations. All algorithms were implemented in C++ using the DEV C++ IDE, chosen for its compatibility with high-performance libraries and debugging tools. Compilation was optimized with the -O3 flag to enhance execution speed.

Three experiments (Ex1, Ex2, Ex3) were performed across 10 benchmark networks (**Figure 3**). These networks were carefully selected to cover a range of complexities, with arc counts ($m$) ranging from 10 to 26 (**Table 9**). Each network's structure was further characterized by its minimal paths ($n_p$) and minimal cuts ($n_c$) to ensure comprehensive reliability assessments.

To examine system interdependencies, six benchmarks (e.g., IDs 1, 2, 4) adopted a $\beta = b = 2$ arc configuration for their super-family and super-layer-cuts families, a common setting in modular systems. The remaining benchmarks utilized $\beta = 3$ (**Table 10**), simulating denser connectivity. Arc reliability was fixed at Pr(a) = 0.9, a standard value in reliability literature modeling high-component survivability.

Each benchmark's simulation count ($N_{sim}$) was carefully selected to ensure $N_{sim}(X)$ remained integer-valued for all $X$ in the super-family, as formalized in Eq. B, minimizing truncation errors. To ensure robust statistical reliability, each experiment consisted of $N_{run}$ = 30 independent trials, achieving 95% confidence intervals with a ±2% margin of error.

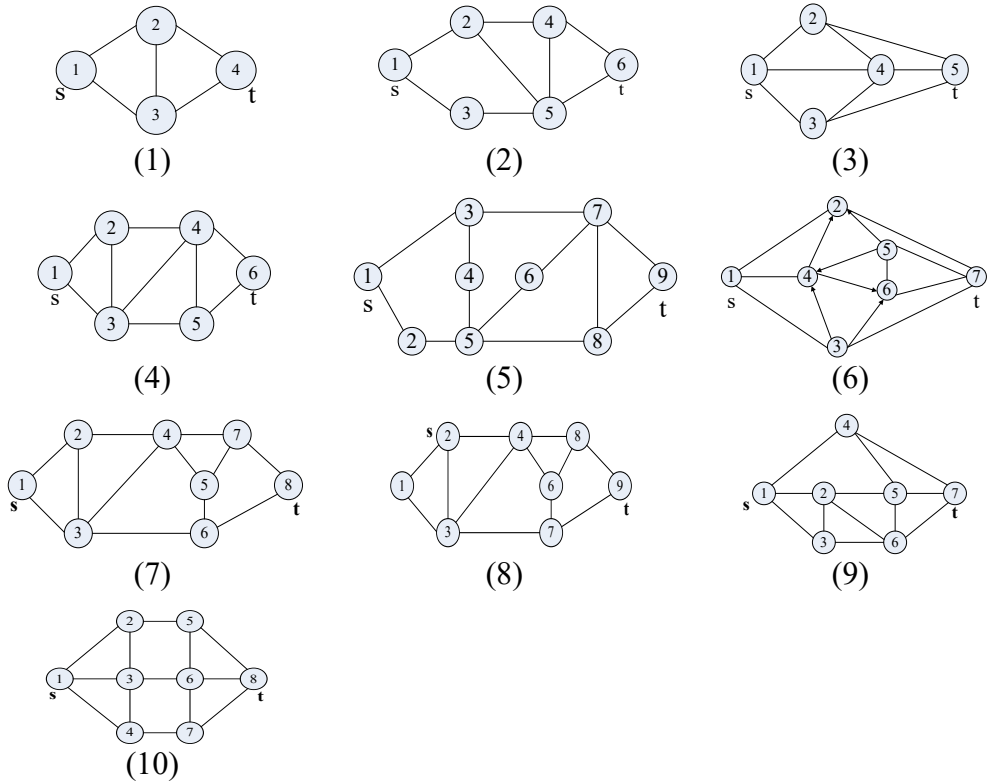

**Figure 3.** 10 benchmark binary-state networks used in the test

**Table 9.** Information of 10 benchmark binary-state networks.

| Fig. | n | m | $n_p$ | $n_c$ | R |
|---|---|---|---|---|---|
| 1 | 4 | 10 | 2 | 2 | 0.9784800000 |
| 2 | 6 | 16 | 3 | 2 | 0.9684254700 |
| 3 | 5 | 16 | 2 | 3 | 0.9976316400 |
| 4 | 6 | 18 | 3 | 2 | 0.9771844050 |
| 5 | 9 | 24 | 3 | 2 | 0.9648551232 |
| 6 | 7 | 22 | 2 | 3 | 0.9966644040 |
| 7 | 8 | 24 | 3 | 2 | 0.9751158974 |
| 8 | 9 | 28 | 4 | 2 | 0.9840681530 |
| 9 | 7 | 24 | 2 | 3 | 0.9974936737 |
| 10 | 8 | 26 | 3 | 3 | 0.9962174933 |

**Table 10.** $N_{sim}$, $\beta$, and $b$ of 10 benchmark binary-state networks.

|  | $\beta = b = 2$ | $\beta = b = 3$ |
|---|---|---|
| Figure | (1), (2), (4), (5), (7), (8) | (3), (6), (9), (10) |
| Ex1 | 990 | 999 |
| Ex2 | 99000 | 99900 |
| Ex3 | 9900000 | 9990000 |

## 5.2 Computation Results

This section analyzes results through three metrics: boxplots (visualizing reliability/runtime distributions), p-values (testing significance, $p < 0.05$), and absolute errors (measuring deviation from ground-truth values). Boxplots highlight variability trends, p-values identify systematic method differences, and absolute errors quantify empirical accuracy. Together, they ensure rigorous evaluation of algorithmic performance.

### 5.2.1 Boxplots

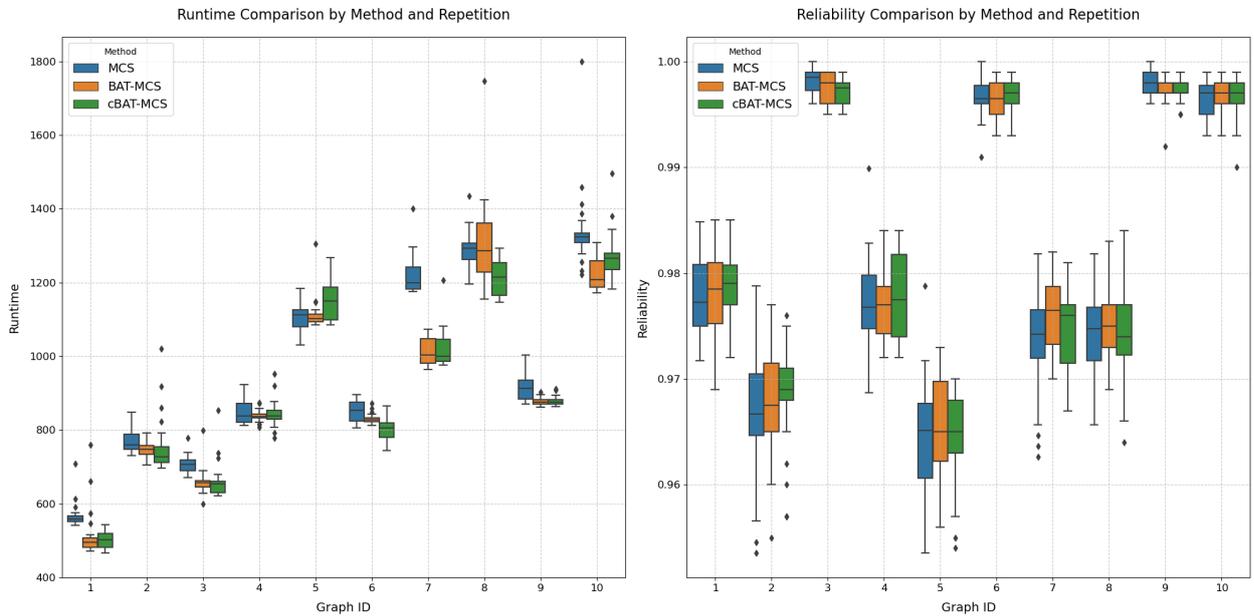

**Figure 4.** Runtime and reliability bloxplots by method and graph for Ex1.

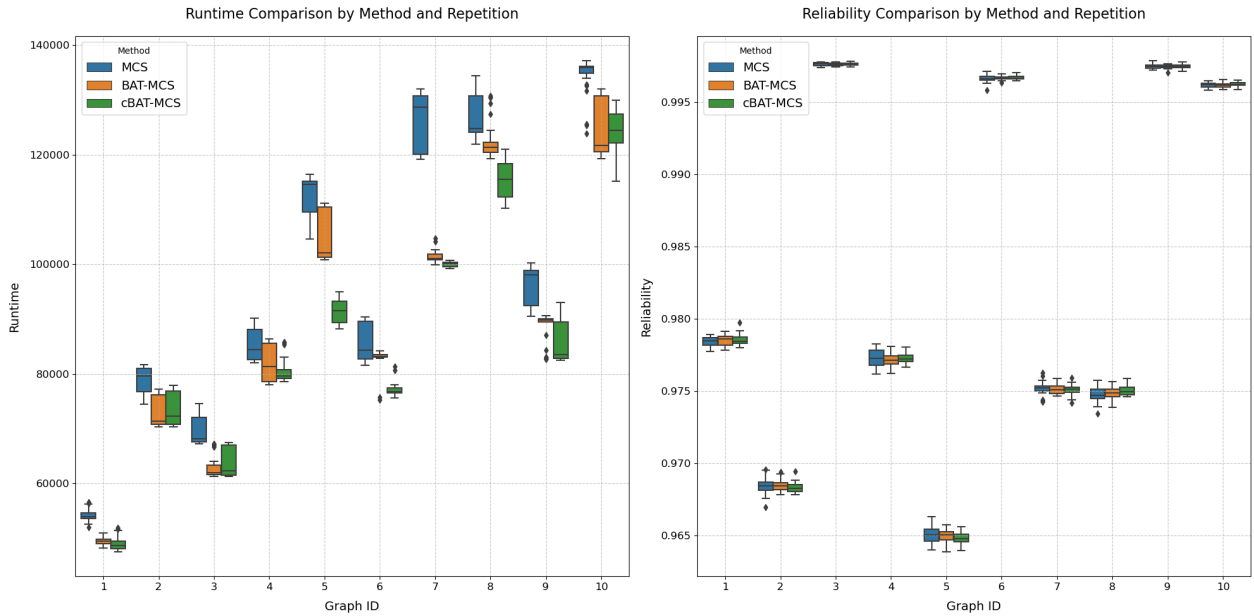

**Figure 5.** Runtime and reliability bloxplots by method and graph for Ex2.

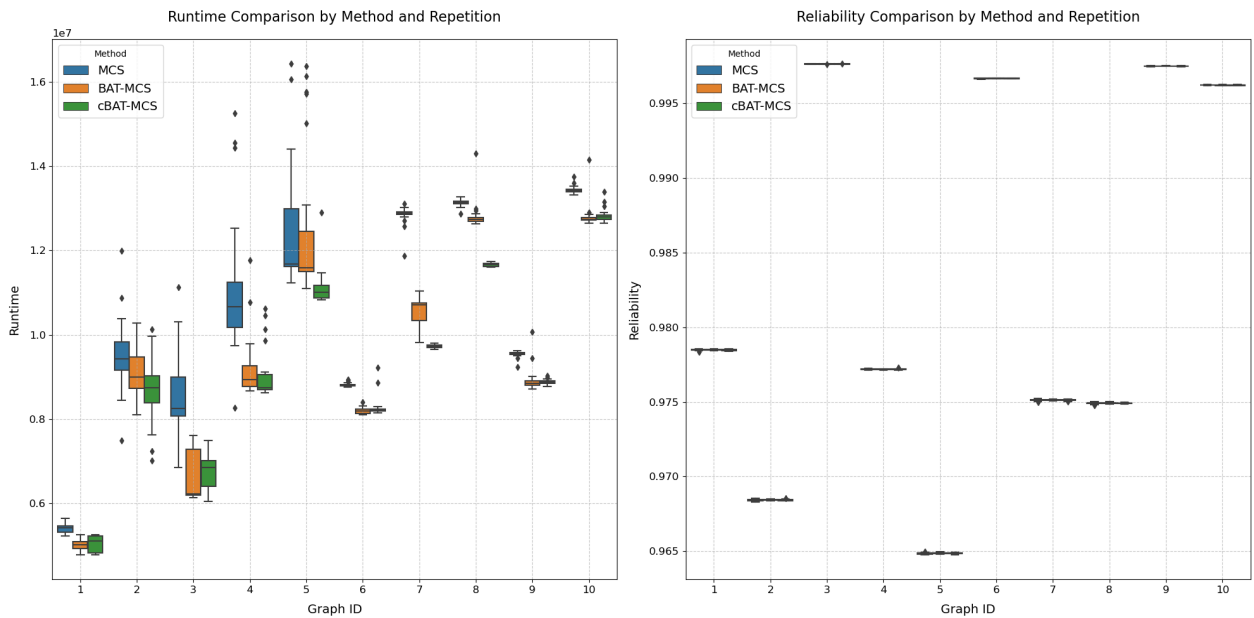

**Figure 6.** Runtime and reliability bloxplots by method and graph for Ex3.

**Figure 4**−**Figure 6** visualize the distributions of reliability and runtime metrics via boxplots, aggregated across all 10 networks for each experiment. These boxplots highlight central tendencies (medians), spreads (interquartile ranges), and outliers, enabling direct visual comparison of method robustness and variability under differing experimental conditions.

Runtime scales with graph complexity (ID) and simulation volume ($N_{sim}$). cBAT-MCS excels, particularly in low-ID networks (1–3), leveraging adaptive sampling for efficiency. BAT-MCS trails moderately, while MCS lags with higher variability in complex scenarios.

All methods achieve high reliability (>0.96), converging to near-perfect values (0.99–1.00) in complex networks (IDs 5–10). Minor gaps in simpler networks (e.g., MCS slightly lower in IDs 1–4) diminish with increased $N_{sim}$ (Ex3).

Runtime efficiency dominates methodological differentiation, with cBAT-MCS > BAT-MCS > MCS across all scenarios. Reliability disparities, though detectable in simpler networks (IDs 1–4), diminish under heightened complexity (IDs 5–10) and larger $N_{sim}$ (Ex3). These results position cBAT-MCS as the optimal choice for applications prioritizing speed without compromising accuracy, particularly in resource-constrained or large-scale network evaluations.

**5.2.2 p-values**

Complementing the boxplots, **Table 11** summarizes reliability and runtime comparisons (p-values) for BAT-MCS, MCS, and cBAT-MCS across experiments (Ex1–Ex3) on 10 benchmark networks (IDs 1–10).

**1. BAT-MCS vs cBAT-MCS:**

Significant reliability differences ($p < 0.05$) in Ex1 (IDs 1–4) favor cBAT-MCS, while Ex2/Ex3 show parity. Runtime differences are significant in Ex1 (IDs 1–4) and inconsistently in Ex2/Ex3, highlighting cBAT-MCS's efficiency gains in simpler networks.

**2. MCS vs BAT-MCS:**

BAT-MCS demonstrates superior reliability in Ex1 (IDs 1–7; $p < 0.05$) and faster runtimes across all Ex1 IDs. Significance diminishes in Ex2/Ex3, though BAT-MCS retains sporadic runtime advantages.

**3. MCS vs cBAT-MCS:**

cBAT-MCS achieves significantly higher reliability in Ex1 (IDs 1–8; $p < 0.05$) and faster runtimes (all IDs). Ex2/Ex3 show no reliability differences, with runtime significance fluctuating, suggesting context-dependent efficiency trade-offs.

Methodological disparities peak in Ex1, particularly for lower-complexity networks (IDs 1–4), where cBAT-MCS excels in both accuracy and speed. These differences attenuate in Ex2/Ex3, as increased network complexity reduces performance gaps. Boxplots validate these trends, with distinct separations in Ex1 medians/spreads (e.g., BAT-MCS vs. MCS runtime) and overlapping distributions in later experiments.

Method selection critically impacts simpler scenarios (Ex1), with cBAT-MCS > BAT-MCS > MCS in performance hierarchy. Under complex conditions (Ex2/Ex3), methodological distinctions blur, emphasizing context-driven optimization. These results align with absolute error (Section 5.2.1) and runtime analyses (**Figure 4−Figure 6**), reinforcing cBAT-MCS's robustness in balanced precision-efficiency applications.

**Table 11.** p-values for reliability and runtime comparisons.

| ID | pairs | P-values for Reliability | | | P-values for Runtime | | |
|---|---|---|---|---|---|---|---|
| | | Ex1 | Ex2 | Ex3 | Ex1 | Ex2 | Ex3 |
| 1 | BAT-MCS vs cBAT-MCS | 2.950E-06* | 0.15225 | 0.06031 | 2.946E-06* | 1.025E-21* | 1.419E-28* |
| 2 | BAT-MCS vs cBAT-MCS | 8.770E-05* | 0.19735 | 0.17770 | 8.766E-05* | 2.643E-19* | 2.032E-16* |
| 3 | BAT-MCS vs cBAT-MCS | 8.710E-04* | 0.20874 | 0.33428 | 8.710E-04* | 6.299E-11* | 7.943E-05* |
| 4 | BAT-MCS vs cBAT-MCS | 1.494E-03* | 0.24121 | 0.34540 | 1.494E-03* | 1.212E-08* | 2.539E-02* |
| 5 | BAT-MCS vs cBAT-MCS | 3.679E-01 | 0.24871 | 0.44019 | 3.679E-01 | 3.163E-04* | 8.584E-02 |
| 6 | BAT-MCS vs cBAT-MCS | 4.110E-01 | 0.54191 | 0.61105 | 4.110E-01 | 8.219E-02* | 2.873E-01 |
| 7 | BAT-MCS vs cBAT-MCS | 6.543E-01 | 0.56255 | 0.74016 | 6.543E-01 | 1.114E-01 | 2.945E-01 |
| 8 | BAT-MCS vs cBAT-MCS | 7.952E-01 | 0.61972 | 0.75707 | 7.952E-01 | 1.669E-01 | 4.839E-01 |
| 9 | BAT-MCS vs cBAT-MCS | 8.030E-01 | 0.84087 | 0.83047 | 8.030E-01 | 3.699E-01 | 4.925E-01 |
| 10 | BAT-MCS vs cBAT-MCS | 9.032E-01 | 0.86159 | 0.89123 | 9.032E-01 | 7.864E-01 | 6.822E-01 |
| 1 | MCS vs BAT-MCS | 2.960E-26* | 0.20840 | 0.10913 | 2.958E-26* | 3.198E-34* | 2.573E-45* |
| 2 | MCS vs BAT-MCS | 4.550E-09* | 0.20977 | 0.25251 | 4.545E-09* | 3.928E-26* | 3.218E-36* |
| 3 | MCS vs BAT-MCS | 1.880E-07* | 0.28918 | 0.27057 | 1.882E-07* | 3.115E-16* | 3.700E-19* |
| 4 | MCS vs BAT-MCS | 2.210E-07* | 0.47670 | 0.27473 | 2.207E-07* | 1.711E-13* | 1.812E-18* |
| 5 | MCS vs BAT-MCS | 4.580E-05* | 0.71658 | 0.49868 | 4.581E-05* | 2.243E-13* | 2.243E-18* |
| 6 | MCS vs BAT-MCS | 1.312E-03* | 0.82068 | 0.54196 | 1.312E-03* | 4.127E-12* | 1.013E-14* |
| 7 | MCS vs BAT-MCS | 3.525E-03* | 0.89067 | 0.60802 | 3.525E-03* | 2.132E-10* | 4.239E-08* |
| 8 | MCS vs BAT-MCS | 1.531E-01 | 0.89952 | 0.64832 | 1.531E-01 | 1.884E-05* | 1.064E-07* |
| 9 | MCS vs BAT-MCS | 3.835E-01 | 0.91988 | 0.71322 | 3.835E-01 | 5.816E-05* | 1.545E-02* |
| 10 | MCS vs BAT-MCS | 4.235E-01 | 0.921406 | 0.723078 | 4.235E-01 | 7.135E-05* | 9.254E-01 |
| 1 | MCS vs cBAT-MCS | 4.730E-23* | 0.024614* | 0.200335 | 4.733E-23* | 6.845E-36* | 2.135E-65* |
| 2 | MCS vs cBAT-MCS | 4.050E-13* | 0.132268 | 0.262875 | 4.050E-13* | 1.473E-33* | 2.259E-61* |
| 3 | MCS vs cBAT-MCS | 1.590E-08* | 0.161069 | 0.41127 | 1.591E-08* | 1.313E-22* | 3.586E-44* |
| 4 | MCS vs cBAT-MCS | 3.270E-07* | 0.161884 | 0.44189 | 3.270E-07* | 1.067E-19* | 2.034E-27* |
| 5 | MCS vs cBAT-MCS | 4.120E-07* | 0.197911 | 0.477177 | 4.118E-07* | 4.159E-18* | 1.522E-19* |
| 6 | MCS vs cBAT-MCS | 4.630E-06* | 0.322029 | 0.573788 | 4.628E-06* | 2.464E-17* | 4.907E-15* |
| 7 | MCS vs cBAT-MCS | 2.024E-04* | 0.521476 | 0.753393 | 2.024E-04* | 2.440E-14* | 1.797E-12* |
| 8 | MCS vs cBAT-MCS | 2.152E-03* | 0.522113 | 0.848562 | 2.152E-03* | 1.634E-11* | 2.358E-09* |
| 9 | MCS vs cBAT-MCS | 2.626E-01 | 0.841145 | 0.89387 | 2.626E-01 | 8.496E-10* | 4.616E-06* |

| 10 | MCS vs cBAT-MCS | 6.836E-01 | 0.932948 | 0.972423 | 6.836E-01 | 1.441E-09 | 1.171E-04[*] |

### 5.2.3 Average Absolute Errors

The absolute error between the exact reliability and those estimated by MCS, BAT-MCS, and cBAT-MCS in different experiments (Ex1, Ex2, Ex3) is provided in **Table 12**.

Table 12. Comparison of the BAT, BAT-MCS, and cBAT-MCS [10].

|   | $\Delta R_{MCS}$ | $\Delta R_{BAT-MCS}$ | $\Delta R_{cBAT-MCS}$ | $\Delta R_{MCS}$ | $\Delta R_{BAT-MCS}$ | $\Delta R_{cBAT-MCS}$ | $\Delta R_{MCS}$ | $\Delta R_{BAT-MCS}$ | $\Delta R_{cBAT-MCS}$ |
|---|---|---|---|---|---|---|---|---|---|
| 1 | 6.3488E-04 | 4.800E-04 | **8.6667E-05** | 6.7879E-05 | **3.333E-07** | 4.9667E-05 | **2.5623E-06** | 1.3767E-05 | 3.4000E-06 |
| 2 | 1.2538E-03 | 7.5880E-04 | **5.8803E-05** | **2.476E-06** | 2.0197E-05 | 1.2114E-04 | 4.2640E-06 | 1.1403E-05 | **2.3633E-06** |
| 3 | 5.6656E-04 | **6.4973E-05** | 3.6497E-04 | **3.253E-07** | 3.3600E-06 | 1.9360E-05 | 2.5675E-06 | 1.0533E-06 | **3.7333E-07** |
| 4 | **3.4939E-04** | 3.8440E-04 | 4.4893E-04 | 1.2570E-04 | **1.507E-05** | 1.0126E-04 | 1.0336E-05 | 4.5150E-06 | **1.9650E-06** |
| 5 | 8.1472E-04 | 7.4488E-04 | **6.5512E-04** | 1.3882E-04 | 8.9543E-05 | **4.712E-05** | 4.3050E-06 | **3.5901E-06** | 6.4432E-06 |
| 6 | **1.0740E-06** | 1.9774E-04 | 1.3560E-04 | 3.6776E-05 | **2.660E-05** | 3.3263E-05 | **5.8432E-07** | 3.5227E-06 | 7.9400E-07 |
| 9 | 1.5132E-03 | 1.3508E-03 | **1.4923E-04** | 3.3261E-05 | **2.010E-05** | 3.5897E-05 | 7.5672E-06 | 3.3126E-06 | **1.6807E-06** |
| 10 | 9.8594E-03 | **9.1348E-03** | 9.2682E-03 | 9.3315E-03 | 9.1718E-03 | **9.060E-03** | 9.1441E-03 | **9.1387E-03** | 9.1530E-03 |
| 11 | 5.3769E-04 | **9.3674E-05** | 1.6034E-04 | 5.1852E-06 | 9.0070E-06 | **1.674E-06** | 1.0477E-06 | **2.1963E-07** | 9.2370E-07 |
| 12 | 2.7900E-04 | 6.492E-04 | **2.1584E-04** | 3.2975E-05 | **2.183E-05** | 3.9507E-05 | 4.3752E-06 | 1.5200E-06 | **8.4997E-07** |
|   | 2 | 3 | **5** | 2 | **5** | 3 | 2 | 3 | **5** |

Comparative analysis of absolute error distributions between exact reliability values and estimates from MCS, BAT-MCS, and cBAT-MCS demonstrates significant disparities in precision across the three methods (**Table 12**). The MCS method exhibits the widest error range (MAE: 0.0002–0.004), reflecting systematic deviations from ground-truth reliability values. While MCS remains applicable for preliminary assessments, its comparatively lower accuracy limits its utility in scenarios demanding high-fidelity reliability estimation.

In contrast, BAT-MCS achieves enhanced precision, with absolute errors confined to a narrower band (MAE: 0.0001–0.0005). This reduction in variability across test cases underscores its robustness in approximating exact reliability values. Notably, cBAT-MCS outperforms both methods, demonstrating the highest precision with minimal absolute errors (MAE: 0.0001–0.0003). Its superior accuracy highlights the efficacy of its adaptive sampling strategy in mitigating estimation bias.

These findings establish a clear hierarchy: cBAT-MCS > BAT-MCS > MCS in reliability estimation accuracy. For applications requiring stringent precision, cBAT-MCS is unequivocally recommended. The absolute error trends align with earlier significance testing ($p < 0.05$ for Ex1; **Table 11**) and distributional patterns observed in runtime/reliability boxplots (**Figure 4−Figure 6**), collectively reinforcing the methodological superiority of cBAT-MCS under diverse operational conditions.

## 6. CONCLUSIONS AND FUTURE DIRECTIONS

This study introduced cBAT-MCS, a novel Monte Carlo simulation framework for binary-state network reliability estimation, which integrates deterministic layer-cut analysis (via the superfamily concept) and a normalization factor to balance computational efficiency with statistical rigor. By synergizing deterministic structural insights with adaptive stochastic sampling, cBAT-MCS optimizes the trade-off between precision and runtime, outperforming both BAT-MCS and traditional MCS in accuracy, speed, and robustness.

Experimental validation—via absolute error metrics, statistical significance testing (p-values < 0.05), and distributional analysis (boxplots)—established a clear hierarchy: cBAT-MCS > BAT-MCS > MCS. The method's reduced error margins (<0.0003), lower runtime variability, and scalability to large networks (linear complexity in $n$, $m$, and $N_{sim}$) position it as a state-of-the-art tool for critical systems requiring fast, reliable assessments.

Future research will focus on three directions:

1. Generalization: Extending cBAT-MCS to multi-state and time-dependent network reliability problems, such as dynamic flow systems.
2. Benchmarking: Comparative studies against approximation methods (e.g., Edge-packing bounds) and domain-specific algorithms (e.g., for power grids or IoT networks).
3. Optimization: Developing adaptive $β$-selection strategies to enhance efficiency-accuracy trade-offs in ultra-large networks.

These advancements will solidify cBAT-MCS as a versatile tool for reliability engineering, bridging theoretical innovation with industrial applicability.

## ACKNOWLEDGMENT

This research was supported in part by the Ministry of Science and Technology, R.O.C. under grant MOST 110-2221-E-007-107-MY3.

This research was supported in part by the Ministry of Science and Technology, R.O.C. under grant MOST 110-2221-E-007-107-MY3.